# A single-cell RNA expression map of coronavirus receptors and associated factors in developing human embryos

S. Colaco†[1], K. Chhabria†[1] D. Singh[1], A. Bhide[1], N. Singh[1], A. Singh[1,2], A. Husein[1], A. Mishra[1], R. Sharma[1], N. Ashary[1], D. Modi*

[1]Molecular and Cellular Biology Laboratory, ICMR-National Institute for Research in Reproductive Health, Indian Council of Medical Research (ICMR), JM Street, Parel, Mumbai 400012, India

[2]MGM Institute of Health Sciences, MGM Educational Campus, Sector 1, Kamothe, Navi Mumbai-410209, India

Corresponding author: deepaknmodi@yahoo.com, modid@nirrh.res.in

† equal contribution

**Running title: Coronavirus receptors and associated factors in human embryos**

**Abstract**

To predict if developing human embryos are permissive to coronaviruses, we analyzed publicly available single cell RNA-seq datasets of zygotes, 4-cell, 8-cell, morula, inner cell mass, epiblast, primitive endoderm and trophectoderm for the coronavirus receptors (*ACE2, BSG, DPP4 and ANPEP*), the Spike protein cleavage enzymes (*TMPRSS2, CTSL*). We also analyzed the presence of host genes involved in viral replication, the endosomal sorting complexes required for transport (ESCRT) and SARS-Cov-2 interactions. The results reveal that *ACE2, BSG, DPP4 and ANPEP* are expressed in the cells of the zygote, to blastocyst including the trophectodermal lineage. *ACE2, TMPRSS,* BSG and *CTSL* are co-transcribed in a proportion of epiblast cells and most cells of the trophectoderm. The embryonic and trophectodermal cells also express genes for proteins ESCRT, viral replication and those that interact with SARS-CoV-2. We identified 1985 genes in epiblast and 1452 genes in the trophectoderm that are enriched in the *ACE2* and *TMPRSS2* co-expressing cells; 216 genes of these are common in both the cell types. These genes are associated with lipid metabolism, lysosome, peroxisome and oxidative phosphorylation pathways. Together our results suggest that developing human embryos could be permissive to coronavirus entry by both canonical and non-canonical mechanisms and they also express the genes for proteins involved in viral endocytosis and replication. This knowledge will be useful for evidence-based patient management for IVF during the COVID-19 pandemic.

**Introduction**

Infection with the novel coronavirus, Severe Acute Respiratory Syndrome-Coronavirus-2 (SARS-CoV-2) causes Coronavirus Disease 2019 (COVID-19) in humans, (Singhal, 2020) and the virus has spread worldwide and infected millions of individuals including pregnant women. Perinatal infections including with coronaviruses can cause maternal and fetal illness and account for 2-3% of all congenital abnormalities (Silasi et al., 2015; Racicot and Mor, 2017; Di Mascio et al., 2020).

Like most viruses, the coronaviruses enter host cells by receptor binding followed by endocytosis, genome replication, exocytosis, and budding. Human Angiotensin-Converting Enzyme II (ACE2) is the host receptor specific to SARS-CoV-2 (Letko et al., 2020) and the S protein of the virus binds to ACE2 with high affinity (Jagtap et al., 2020; Shang et al., 2020a, 2020b). SARS-CoV-2 does not use other coronavirus receptors viz; alanine aminopeptidase N (ANPEP) and dipeptidyl peptidase 4 (DPP4), which are used by CoV-229E and the Middle East Respiratory Syndrome Coronavirus (MERS-CoV), respectively, for entry into host cells (Li, 2015; Letko et al., 2020). In addition to ACE2, SARS-CoV-2 utilizes Basigin (BSG), also known as CD147, as an alternate receptor for viral entry (Wang et al., 2020). The binding of the virus to the host receptors occurs via the SARS-CoV-2 spike (S) protein that is processed by a range of host proteases (Hoffmann et al., 2020; Shang et al., 2020a) like transmembrane serine protease 2 (TMPRSS2), TMPRSS4, FURIN, and cathepsin L (CTSL). The host receptors and S protein processing enzymes are collectively referred to as SARS coronavirus-associated receptors and factors (SCARFs), and their co-expression in the tissues is the determinant of susceptibility to SARS-CoV-2 infection (Singh et al., 2020b). SCARFs are expressed in a variety of human tissues such as the airway and alveolar epithelium, oral and nasal mucosa, nasopharynx, stomach, small intestine, colon, skin, lymph nodes, thymus, bone marrow, spleen, liver, kidney, and placenta (Lukassen et al., 2020; Qi et al., 2020; Seow et al., 2020; Singh et al., 2020b; Xu et al., 2020).

Epidemiologic and clinical evidence suggests that pregnant women are at higher risk of severe illness and mortality from viral infections, which can lead to several adverse outcomes such as spontaneous abortion, intrauterine fetal deaths, still-births, and mother-to-child transmission resulting in congenital viral syndromes (Silasi et al., 2015; Racicot and Mor, 2017; Adams Waldorf et al., 2018). In pregnant women, COVID-19 is associated with severe pregnancy complications such as preterm labor and premature rupture of membranes (Gajbhiye et al., 2020; Knight et al., 2020). A small but significant proportion of women with COVID-19 experience still-births or intrauterine fetal death, and there is evidence of vertical transmission of the virus through the placenta (Bahadur et al., 2020;

Gajbhiye et al., 2020; Hosier et al., 2020; Knight et al., 2020; Verma et al., 2020; Vivanti et al., 2020). Indeed, SCARFs are expressed in the trophoblast cells (Ashray et al., 2020; Li et al., 2020a; Singh et al., 2020b) suggesting that the placenta may be permissive to SARS-CoV-2 infection.

Beyond the adult tissues, viruses can also infect gametes and developing embryos (Racicot and Mor, 2017). It has been recently shown that Zika virus can infect and propagate in the trophectodermal cells of preimplantation human, rhesus monkey, and mouse embryos, causing death of neural progenitors leading to microcephaly and miscarriage (Tan et al., 2019; Block et al., 2020). Thus, studies on preimplantation mammalian embryos are important because viral infections during early pregnancy can induce embryonic death or abnormal embryonic development (Hardy, 1974; Silasi et al., 2015; Tan et al., 2019; Block et al., 2020). With the high population spread and the observation that a large proportion of individuals with SARS-CoV-2 infection may be asymptomatic, (Bharti et al., 2020; Hao et al., 2020; Kaur et al., 2020; Kissler et al., 2020; Li et al., 2020b) it is likely that many women may have conceived during the course of subclinical infection. Secondly, many assisted reproduction clinics may have handled human gametes and embryos from sub-clinically infected couples or inadvertently exposed them to infected laboratory personnel. Given the fact that SARS-CoV-2 has a high surface stability on various materials (Ong et al., 2020; van Doremalen et al., 2020), there can be many potential routes by which gametes/embryos could be exposed to SARS-CoV-2. However, whether the human gametes or early embryos are permissive to SARS-CoV-2 or other coronaviruses remains unknown.

As a preamble towards our understanding of the potential effects of SARS-CoV-2 in early pregnancy, we analysed publicly available single-cell RNA-Seq datasets of human embryos to determine the expression of SCARFs in the zygote to late-blastocyst-stage human embryos. In addition, we determined whether endocytotic and viral replication machinery are also expressed by the cells of early embryos.

**Materials and Methods**

To understand the expression of SCARFs in various cells of the embryo (zygote to late blastocysts), single-cell RNA-Seq datasets (Petropoulos et al., 2016; Stirparo et al., 2018) were analysed. The dataset by Stirparo *et al.* was chosen as it contained uniformly analysed samples from the three published studies (Yan et al., 2013; Blakeley et al., 2015; Petropoulos et al., 2016) and recapitulated the known lineage markers of zygote to late blastocyst embryos. It allows an in-depth exploration of the transcriptome of multiple cells in the inner cell mass (ICM), epiblast, and primitive endoderm.

The dataset from Petropoulos *et al.,* 2016 (accession number: E-MTAB-3929 allowed in-depth exploration of the largest numbers of trophectodermal cells from day 5 to day 7 blastocysts of *in vitro* fertilization. Table 2 and Table 3 give the number of embryos and the total number of cells tested at each time point of embryonic development.

The percentage of *ACE2*, *BSG*, *TMPRSS2*, *CTSL*, *DPP4*, and *ANPEP* expressing cells were obtained for each cell at each stage; the percentage of cells co-expressing *ACE2* and *BSG*, *ACE2* and *TMPRSS2*, *ACE2* and *CTSL*, and *BSG* and *CTSL* were also estimated for each cell type. The list of genes representing protein products involved in the Endosomal Sorting Complex Required for Transport (ESCRT I, II, and III) and SARS virus replication in host cells were obtained (de Wilde et al., 2018; Ahmed et al., 2019), and their mRNA levels in cells of the zygote to blastocyst (including trophectoderm) were extracted. In addition, the data containing mRNA levels of 332 proteins that physically interact with SARS-CoV-2 (Gordon et al., 2020) were extracted.

We aimed to identify the gene signatures associated with *ACE2+TMPRSS2+* cells in the developing embryos. Pseudo-bulk analysis of single-cell data of the epiblast and trophectodermal lineage on day 7 was carried out for *ACE2*- and *TMPRSS2*-positive (*ACE2+TMPRSS2+*) and *ACE2*- and *TMPRSS2*-negative (*ACE2–TMPRSS2–*) cells. The epiblast cells and day 7 trophectoderm cells were chosen as they had a reasonable proportion of both *ACE2+TMPRSS2+* co-expressing cells versus those that did not express either. This allowed us to have enough numbers of cells in both groups to reasonably apply statistical testing. For pseudo-bulk analysis, the mean values of all the genes expressed in the *ACE2+TMPRSS2+* and *ACE2–TMPRSS2–* cells were computed. Genes that had a ratio of $\geq$1.5 or $\leq$0.5 were filtered. Welch's t-test was applied and genes that had significantly different mean values (p$\leq$0.05) were filtered.

The Gene Ontologies (GO) associated with the differentially expressed genes in *ACE2*- and *TMPRSS2*-positive cells were obtained using gProfiler (https://biit.cs.ut.ee/gprofiler/gost). Categories with Benjamini-Hochberg False Discovery Rate [FDR] <0.05 were considered as enriched. The terms were selected based on negative_log10_of_adjusted_p_value. Pseudo-bulk and deconvoluted single-cell data were visualized using Morpheus (https://software.broadinstitute.org/morpheus) and R Studio version 3.6.2 with the Seurat and ggplot2 packages (RStudio Team (2020). RStudio: Integrated Development for R. RStudio, PBC, Boston, MA http://www.rstudio.com/).

# Results

## mRNA levels of coronavirus-associated genes *ACE2*, *TMPRSS2*, *BSG*, *CTSL*, *DPP4*, and *ANPEP* in early stages of human embryonic development

Table1. summarizes the receptors used by the various coronaviruses known to infect humans. As evident from Fig.1 *ACE2* mRNA was detected in the zygotes, which decreased until the morula stage and subsequently increased in the ICM, epiblast, and the primitive endoderm. Almost all cells of the trophectoderm of day 5 and day 6 embryos expressed *ACE2*, and the numbers of cells and transcript abundance of *ACE2* declined in trophectoderm of day 7 blastocysts.

**Table 1: Summary of host receptors of coronaviruses that infect humans**

| Virus | Human host receptor |
|---|---|
| SARS-CoV-2 | ACE2, CD147 (BSG) |
| SARS-CoV | ACE2, CD147 (BSG) |
| MERS-CoV | DPP4 |
| HCoV-OC43 | 9-O-acetylsialic acids |
| HCoV-HKU1 | 9-O-acetylsialic acids |
| HCoV-229E | ANPEP |
| CoV-NL63 | ACE2 |

*BSG* mRNA was consistently detected in all stages of embryonic development with an increase in expression from the zygote to blastocyst stage (Fig. 1). Almost all the trophectodermal cells of the day 5-7 blastocysts expressed *BSG* mRNA (Fig. 1).

*TMPRSS2* was not detected in the zygotes, and very few *TMPRSS2*-positive cells were detected until the blastocyst stage. The highest mRNA levels of *TMPRSS2* were observed in epiblast cells, but their numbers were lower in the primitive endoderm (Fig. 1). Abundant *TMPRSS2* was detected in most of the trophectodermal cells of day 5-7 blastocysts (Fig. 1).

A low abundance of *CTSL* mRNA was observed in the zygotes and 4- and 8-cell embryos, but its levels increased in all the cells of the morula, ICM, epiblast, and primitive endoderm (Fig. 1). *CTSL* mRNA was expressed in most trophectodermal cells of the day 5-7 embryos (Fig. 1).

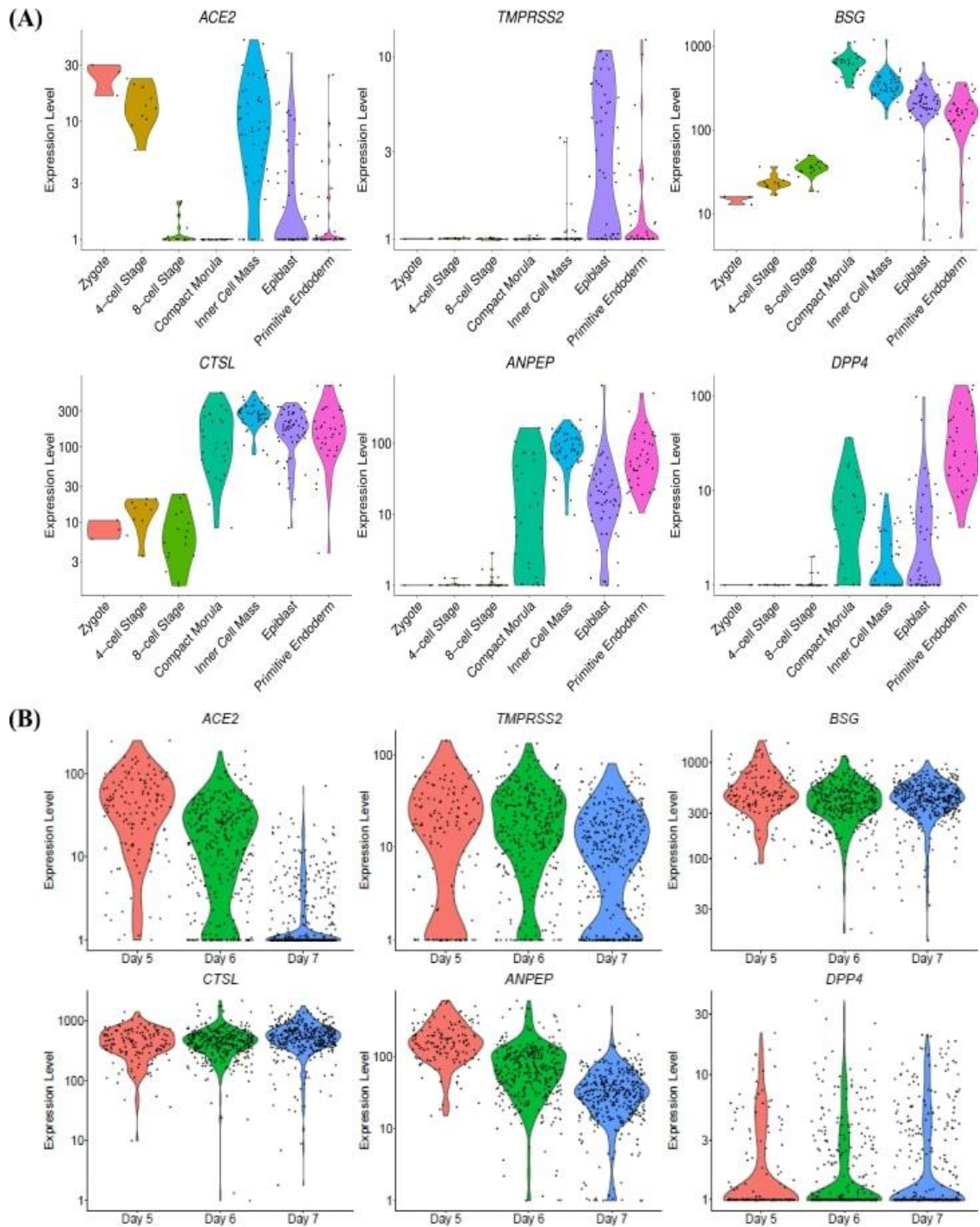

**Fig.1. mRNA expression of coronavirus receptors and spike protein processing enzymes in early human embryos.** (A) X axis represents various stages of development from zygote to blastocysts and Y axis represents FPKM values. (B) X axis represents trophectoderm of day 5 to day 7 and Y axis represents RPKM value. Each dot represents expression of mRNA in single cell. ACE2 and BSG are the receptors for SARS-CoV-2 and SARS-CoV, DPP4 is the receptor for MERS-CoV, and ANPEP is the receptor for HCoV-229E. The viral spike protein is processed by host enzymes TMPRSS2 or CTSL. scRNAseq data for embryonic cells (in A) was extracted from (Stirparo et al.,

2018) while that of trophectodermal cells on different days of embryonic development (in B) was extracted from (Petropoulos et al., 2016).

*DPP4* and *ANPEP* mRNA was not detected in the zygote and 4- and 8-cell embryos; however, abundant *DPP4* and *ANPEP* transcripts were detected in the cells of the morula, ICM, epiblast, and primitive endoderm (Fig.1). Both *DPP4* and *ANPEP* were also abundantly expressed in trophectodermal cells of day 5-7 blastocysts (Fig. 1).

**Co-expression of transcripts of SARS-CoV-2 receptors and processing enzymes in early stages of human embryonic development**

We next determined the numbers of cells at different stages of embryonic development that co-express transcripts of *ACE2* and *TMPRSS2*. Cells of the zygote, 4- and 8-cell, and morula stage embryos did not co-express *ACE2* and *TMPRSS2*. A small proportion of cells in the ICM and primitive endoderm co-expressed *ACE2* and *TMPRSS2*, and the largest proportion (34%) of *ACE2* and *TMPRSS2* co-expressing cells were detected in the epiblast (Fig. 2). In the trophectodermal lineage, the day 5 blastocysts had 77% (110/142) and day 6 blastocysts had 83% (275/331) of cells co-expressing *ACE2* and *TMPRSS2*. The numbers of *ACE2* and *TMPRSS2* double-positive trophectodermal cells declined to 33% (129/388) in day 7 blastocysts (Fig. 3).

We also determined the co-expression of *ACE2* and *CTSL*, *ACE2* and *BSG*, and *BSG* and *CTSL* in single cells of developing human embryos. The results reveal that all the zygotes and all cells of the 4- cell stage co-expressed *ACE2* and *BSG* (Fig. 2). In the ICM, around 86% (37/43) of cells co-expressed *ACE2* and *BSG*; 39% (25/64) of the cells of the epiblast and 39% (11/28) of the cells of the primitive endoderm co-expressed *ACE2* and *BSG* (Table 2). Most trophectodermal cells of day 5 and day 6 blastocysts co-expressed *ACE2* and *BSG* (Fig. 3); on day 7, the number of trophectodermal cells co-expressing *ACE2* and *BSG* declined to 38% (149/388)(Table 3).

**Table 2:** Numbers and percentage of cells expressing *ACE2+TMPRSS2+*, *ACE2+BSG+*, *ACE2+CTSL+*, *BSG+CTSL+* in human embryos at different stages. Data was extracted from single-cell RNA-Seq of the epiblast of developing human embryos (Stirparo et al 2018). The total cells and the numbers of embryos analysed are also given.

|  | *ACE2+ TMPRSS2+* | *ACE2+ BSG+* | *ACE2+ CTSL+* | *BSG+ CTSL+* | Total Cells Analysed | Total Embryos Analysed |
|---|---|---|---|---|---|---|
| Zygote | 0 (0%) | 3 (100%) | 3 (100%) | 3 (100%) | 3 | 3 |
| 4-Cell Stage | 0 (0%) | 10 (100%) | 10 (100%) | 10 (100%) | 10 | 3 |
| 8-Cell Stage | 0 (0%) | 5 (31%) | 5 (31%) | 16 (100%) | 16 | 3 |
| Morula | 0 (0%) | 3 (10%) | 3 (10%) | 29 (100%) | 29 | 3 |
| Inner Cell Mass | 5 (12%) | 37 (86%) | 37 (86%) | 43 (100%) | 43 | 3 |
| Epiblast | 22 (34%) | 25 (39%) | 25 (39%) | 64 (100%) | 64 | 12 |
| Primitive Endoderm | 6 (21%) | 11 (39%) | 11 (39%) | 28 (100%) | 28 | 12 |

**Table 3:** Numbers and percentages of cells expressing *ACE2+TMPRSS2+*, *ACE2+BSG+*, *ACE2+CTSL+*, and *BSG+CTSL+* in trophectoderm of human embryos at different stages, as well as total numbers of cells and embryos analysed. Data were extracted from single-cell RNA-Seq of trophectoderm (Petropoulos et al., 2016, accession no. E-MTAB-3929).

|  | *ACE2+ TMPRSS2+* | *ACE2+ BSG+* | *ACE2+ CTSL+* | *BSG+ CTSL+* | Total Cells Analysed | Total Embryos Analysed |
|---|---|---|---|---|---|---|
| Day 5 | 110 (77%) | 140 (99%) | 140 (99%) | 142 (100%) | 142 | 18 |
| Day 6 | 275 (83%) | 290 (88%) | 290 (88%) | 330 (100%) | 331 | 17 |
| Day 7 | 129 (33%) | 149 (38%) | 149 (38%) | 388 (100%) | 388 | 17 |

*ACE2* and *CTSL* were co-expressed in the zygotes and all the cells of 4-cell stage embryos, 31% of cells in the 8-cell stage, and 10% of the cells in the morula. Around 86% (37/43) of ICM, 39% (25/64) of epiblast, and 39% (11/28) of primitive endoderm cells co-expressed *ACE2* and *CTSL* (Fig. 2). In the trophectoderm, 99% (140/142), 88% (290/331), and 38% (149/388) of trophoblasts from day 5, 6, and 7 blastocysts co-expressed *ACE2* and *CTSL* (Fig. 3).

*BSG* and *CTSL* co-expressing cells were observed at all stages of development. All the cells of the ICM (43/43), epiblast (64/64), and primitive endoderm (28/28) co-expressed *BSG* and *CTSL* (Fig. 2). In the trophectodermal lineage, 100% (142/142) of day 5, 330/331 day 6, and 100% (388/388) of day 7 blastocysts co-expressed *BSG* and *CTSL* (Fig. 3).

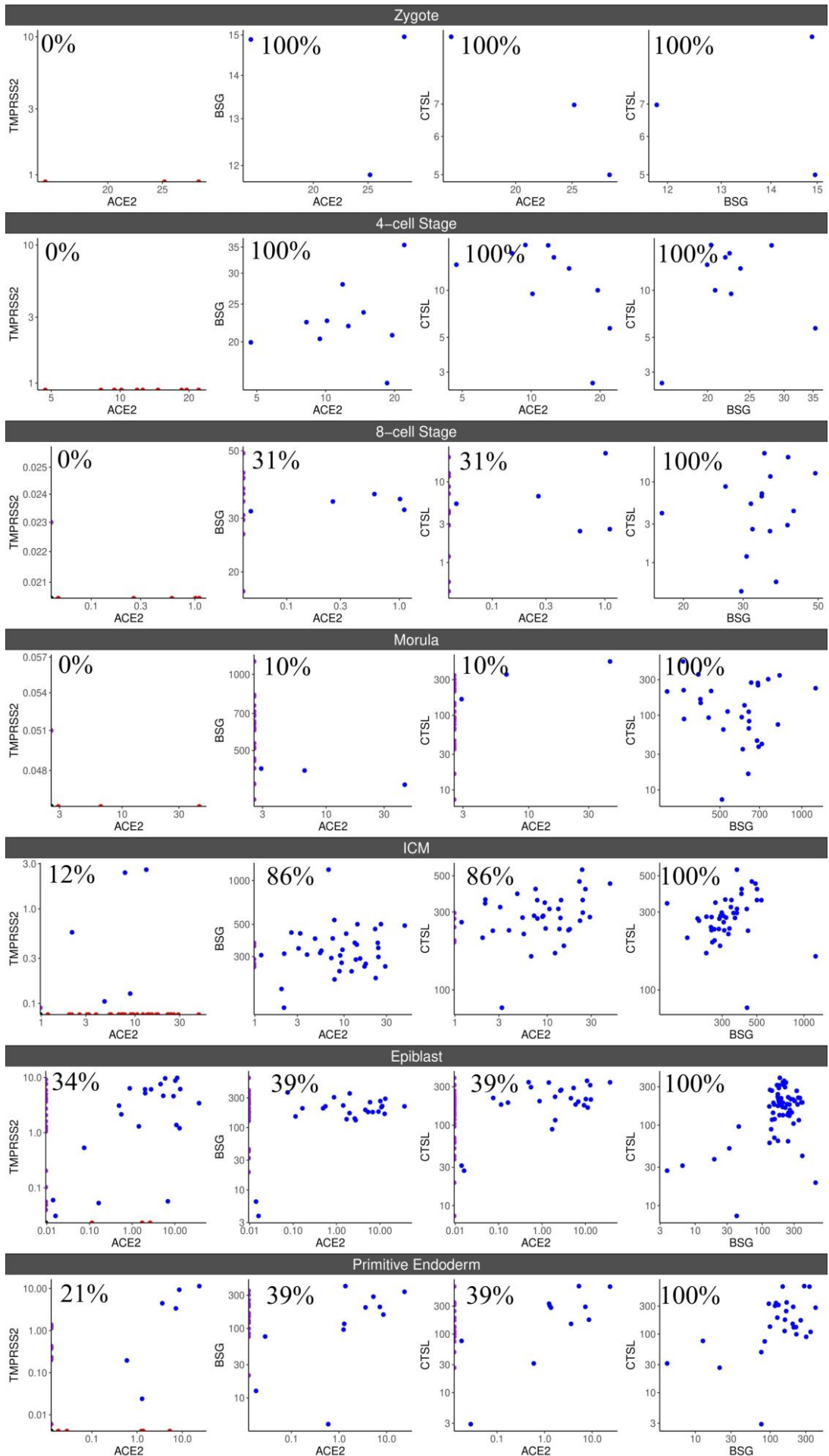

**Fig.2. Co-expression of mRNA of SARS-CoV-2 receptors and spike protein processing enzymes during early stages of embryonic development.**

Co-expression of *ACE2* and *TMPRSS2*, *ACE2* and *BSG, ACE2* and *CTSL*, and *BSG and CTSL* in single cells of human embryos at zygote, 4 cell, 8 cell, compact morula, early inner cell mass (ICM) epiblast and primitive endoderm. Blue dot represents expression of both the transcripts plotted on X as well Y axis, red dot represents expression of only one transcript plotted on X axis and magenta dot represents expression of only one transcript plotted on Y axis. The numbers in each graph represents percentage (%) of co-expressing cells. Data was extracted from single-cell RNA-Seq of developing human embryos (Stirparo et al., 2018).

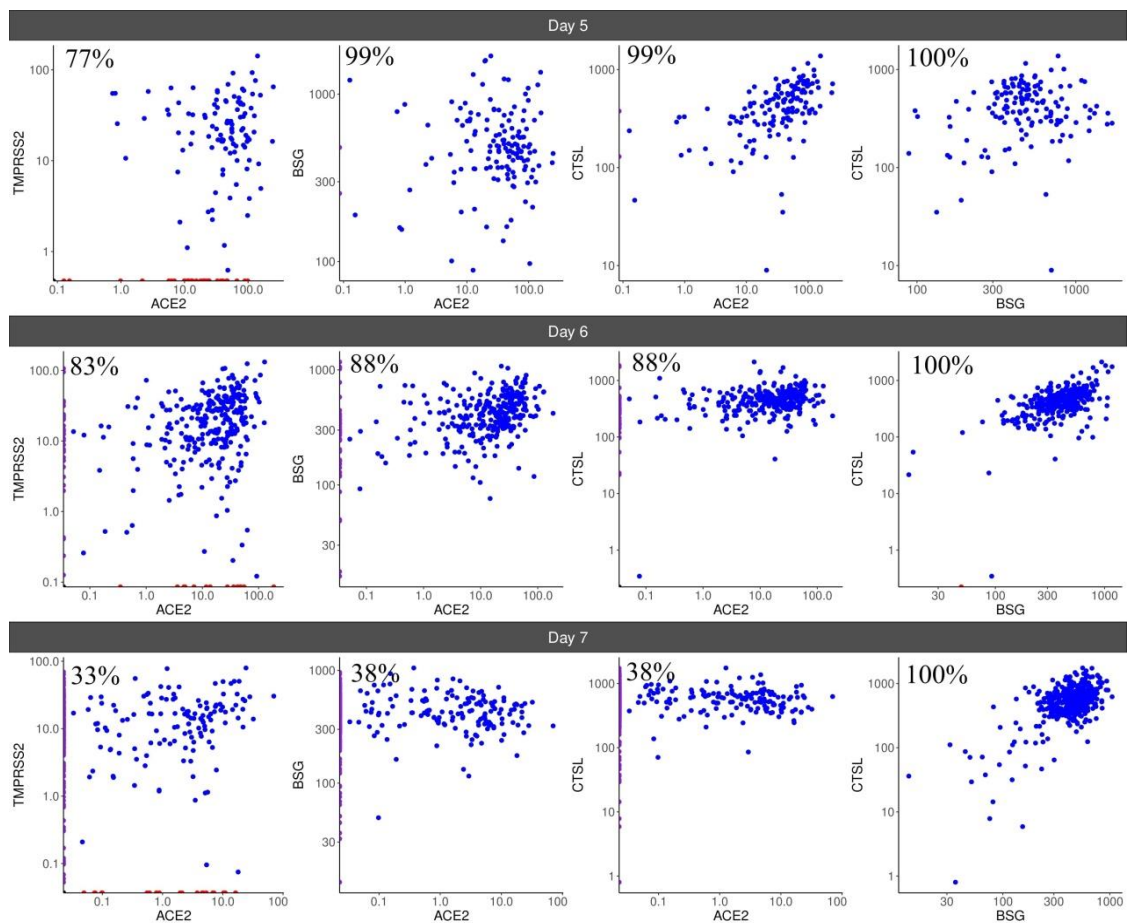

**Fig.3 Co-expression of mRNA of SARS-CoV-2 receptors and spike protein processing enzymes in trophectoderm lineage at day 5 to day 7.** Co-expression of *ACE2* and *TMPRSS2*, *ACE2* and *BSG, ACE2* and *CTSL*, and *BSG and CTSL* in trophectoderm of human embryo. Blue dot represents expression of both the transcripts, red and magenta dot represents expression of only one transcript.

The numbers in each graph represents percentage (%) of co-expressing cells. Data was extracted from single-cell RNA-Seq of trophectoderm of day 5-7 blastocysts (Petropoulos et al., 2016).

**mRNA levels of genes involved in viral endocytosis and replication in human embryos at different stages of development.**

We next hypothesized that if the virus can bind to the embryonic cells, the embryos may express the cellular machinery necessary for viral endocytosis and viral replication. Towards this, we analysed mRNA levels of the 33 genes involved in the human ESCRT of viruses. The pseudo-bulk analysis revealed that from the zygote to the morula stage, the mRNA for most of the ESCRT genes is undetected or present in low abundance. However, the transcripts for almost 90% of these genes dramatically increased in the ICM, epiblast, and primitive endoderm. Additionally, almost all the genes were expressed in high abundance in the trophectodermal lineage of day 5-7 embryos (Fig.4).

Coronaviruses enter the cells and utilizes the host cell machinery for replication. We determined the mRNA levels for host genes involved in SARS-CoV replication by pseudo-bulk analysis and observed that the mRNA of genes representing protein products involved in viral replication was not expressed or was present in low abundance from the zygote to morula stages of embryonic development. However, the mRNA levels of most of these genes surged in the ICM, epiblast, and the primitive endoderm, and maximal expression of almost all these genes was observed in the trophectodermal cells of day 5-7 embryos (Fig. 4).

SARS-CoV-2 interacts with 332 proteins in human cells. Transcripts for only a few of these genes were abundant; most were weakly expressed in the zygote, 4- and 8-cell, morula, and ICM stages as evident by pseudo-bulk analysis. However, many cells of the epiblast and primitive endodermal lineage expressed the mRNA for these genes. Almost all the cells of the trophectodermal lineage of day 5-7 blastocysts expressed mRNA encoding for the proteins that interact with SARS-CoV-2 (Fig4).

Analysis of individual cells of the epiblast and day 7 blastocyst, which had proportions of *ACE2* and *TMPRSS2* co-expressing cells (most permissive to SARS-CoV-2 infection), revealed that almost all the genes representing protein products involved in ESCRT, viral replication, and SARS-CoV-2 interaction were expressed with minimal cell-to-cell heterogeneity (Supplementary Fig 1). Similar results were observed for day 5 and day 6 blastocysts (not shown).

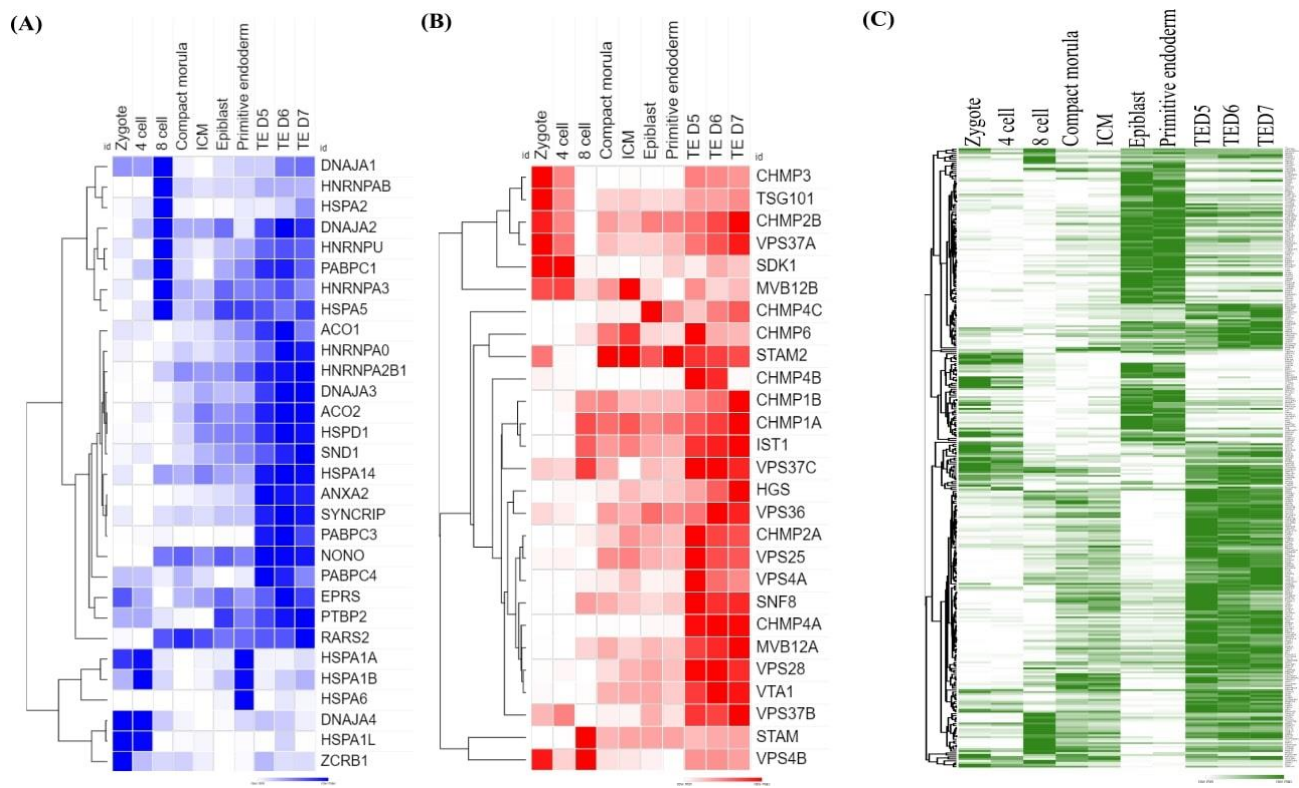

**Fig.4. mRNA levels of genes involved in viral endocytosis and replication in human embryos at different stages of development.** Pseudo-bulk data for the mRNA levels of genes involved in (A) Endosomal Complexes Required for Transport (ESCRT), (B) viral replication and (C) SARS-CoV-2 interactions. All data were extracted from single-cell RNA-Seq of developing human embryos (Stirparo et al., 2018) and from trophectoderm cells from (Petropoulos et al., 2016) In all heat maps, each row depicts a gene and each column depicts a stage of embryonic development. Data is presented in relative color scale.

**Differentially expressed genes and the biological processes enriched in *ACE2* and *TMPRSS2* co-expressing cells of the epiblast and trophectodermal cells of day 7 blastocysts.**

In the epiblast, 34% of cells co-expressed *ACE2* and *TMPRSS2* while the others did not express either *ACE2* or *TMPRSS2* (Table 2). By pseudo-bulk analysis, we identified 1985 genes that were differentially expressed (fold change >1.5 or <0.5) with statistical significance (p-value <0.05) between *ACE2-* and *TMPRSS2*-positive cells and *ACE2-* and *TMPRSS2*-negative cells of the epiblast. Fig. 5 depicts the deconvolution of these 1985 differentially expressed genes in single cells of the epiblast. As compared to *ACE2-* and *TMPRSS2*-negative cells, nearly 67% (1323/1985) of genes were present at a higher abundance in the *ACE2-* and *TMPRSS2*-positive cells of the epiblast. While there was cell-to-cell heterogeneity in expression of the differentially abundant genes in the *ACE2–TMPRSS2–* cells, many of these genes were uniformly expressed in most *ACE2+ TMPRSS2+* cells (Fig 5).

In the trophectodermal lineage of day 7 blastocysts, 71 cells were negative for both *ACE2* and *TMPRSS2* transcripts whereas 97 cells expressed both the transcripts above the cut-off. Pseudo-bulk analysis identified 1452 differentially abundant genes between *ACE2+TMPRSS2+* and *ACE2–TMPRSS2–* cells. Of these, 1212 genes were present in higher abundance and 240 genes were present at lower abundance in the *ACE2+TMPRSS2+* cells (Supplementary Table 4). The expression of the majority of these genes was homogeneous in the *ACE2* and *TMPRSS2* co-expressing cells, and expression was highly heterogeneous in the non-expressing cells (Fig. 5).

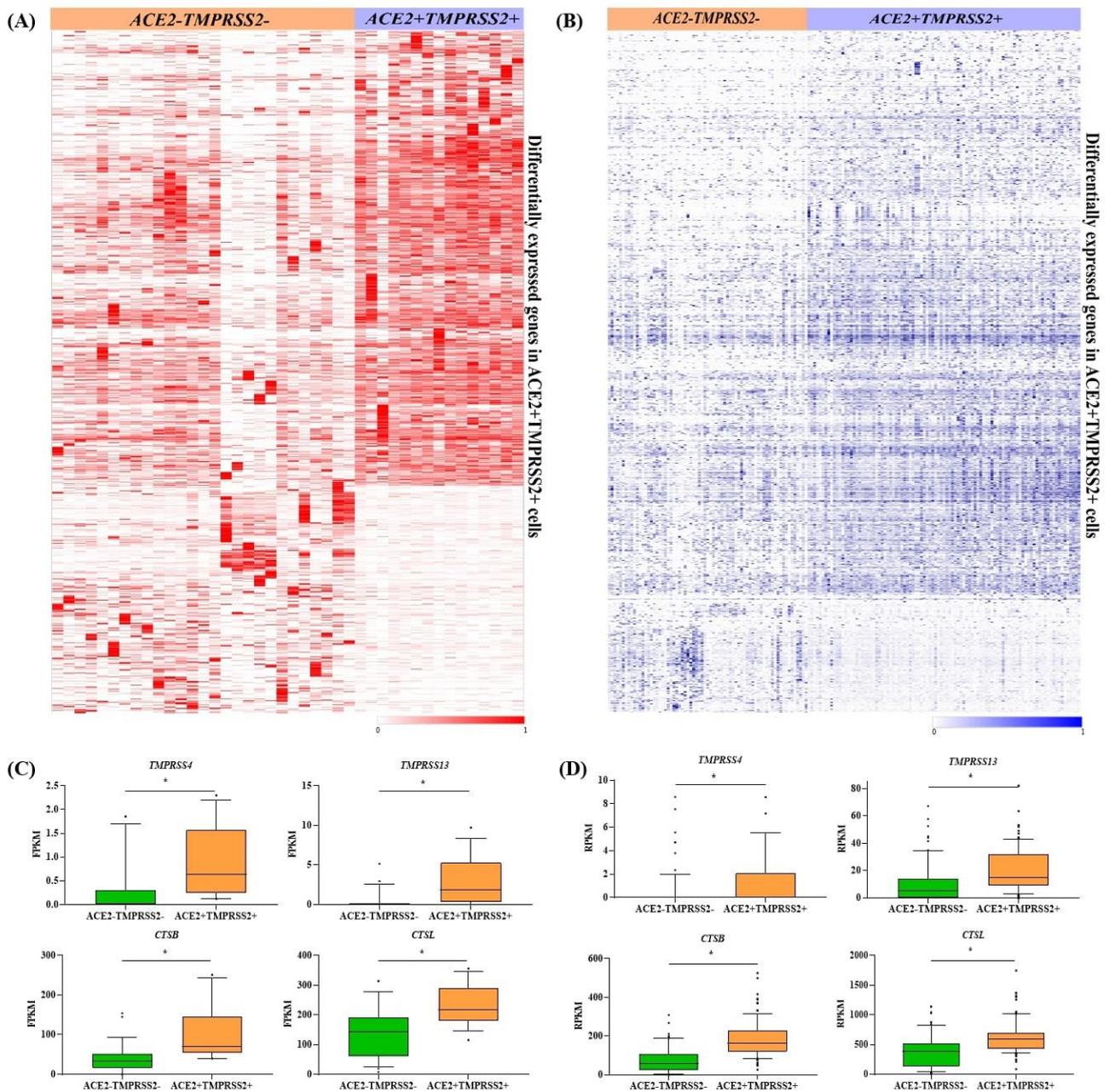

**Fig.5. Differentially expressed genes in *ACE2*- and *TMPRSS2*-positive cells in epiblast and trophectodermal cells of developing human blastocysts.** (A) Heat map of distribution of differentially expressed genes (1985 genes) in *ACE2*- and *TMPRSS2*-negative (*ACE2−-TMPRSS2−*) and *ACE2*- and *TMPRSS2*-positive (*ACE2+-TMPRSS2+*) cells of the epiblast. (B) Heat map of distribution of 1452 differentially expressed genes in *ACE2−-TMPRSS2−* and *ACE2+TMPRSS2+* trophectodermal cells. In A and B rows represent genes and columns represent individual cells, presented on a relative color scale. (C) Expression of *TMPRSS4, TMPRSS13, CTSB* and *CTSL* in *ACE2−TMPRSS2−* and *ACE2+TMPRSS2+* cells of epiblast (D) Expression of *TMPRSS4, TMPRSS13, CTSB* and *CTSL* in *ACE2−TMPRSS2−* and *ACE2+TMPRSS2+* cells of trophectoderm.

Significantly different (p<0.05) mean values are marked as *. Data were extracted from single-cell RNA-Seq of the epiblast of developing human embryos (Stirparo et al., 2018), trophectoderm data was extracted from (Petropoulos et al., 2016).

In both epiblast and trophectodermal cells, the *ACE2+TMPRSS2+* cells were also enriched for other S protein cleavage enzymes. In the epiblast and day 7 trophectoderm, *TMPRSS4, TMPRSS13, CTSB* and *CTSL* transcripts were present in significantly higher abundance in the *ACE2+TMPRSS2+* cells as compared to the cells that did not express either of the two genes (p-value ≤0.05). Many other members of the TMPRSS and Cathepsin family were also enriched in the *ACE2+TMPRSS2+* cells, although many did not reach statistical significance (Supplementary Fig. 2)

Comparative analysis of the differentially abundant transcripts identified 216 genes that were common in both the epiblast and day 7 trophectodermal lineage, while 1768 genes were unique to epiblast and 1236 genes were unique to the trophectodermal lineage (Fig. 6). The list of these genes is given in Supplementary Table 5.

The biological processes associated with the 216 common genes involved sterol, alcohol and cholesterol metabolism and localization; the enriched KEGG pathways included lysosome and steroid biosynthesis (Fig. 6). The genes unique to epiblast cells involved biological processes associated with embryonic development and energy metabolism; enriched KEGG pathways included oxidative phosphorylation and small molecule metabolism (Fig. 6). The genes unique to the trophectoderm lineage involved biological processes associated with glycolipid metabolism, liposaccharide metabolism, and leukocyte activation; enriched KEGG pathways included lysosome, peroxisome, glycan degradation and herpes simplex virus infection (Fig. 6).

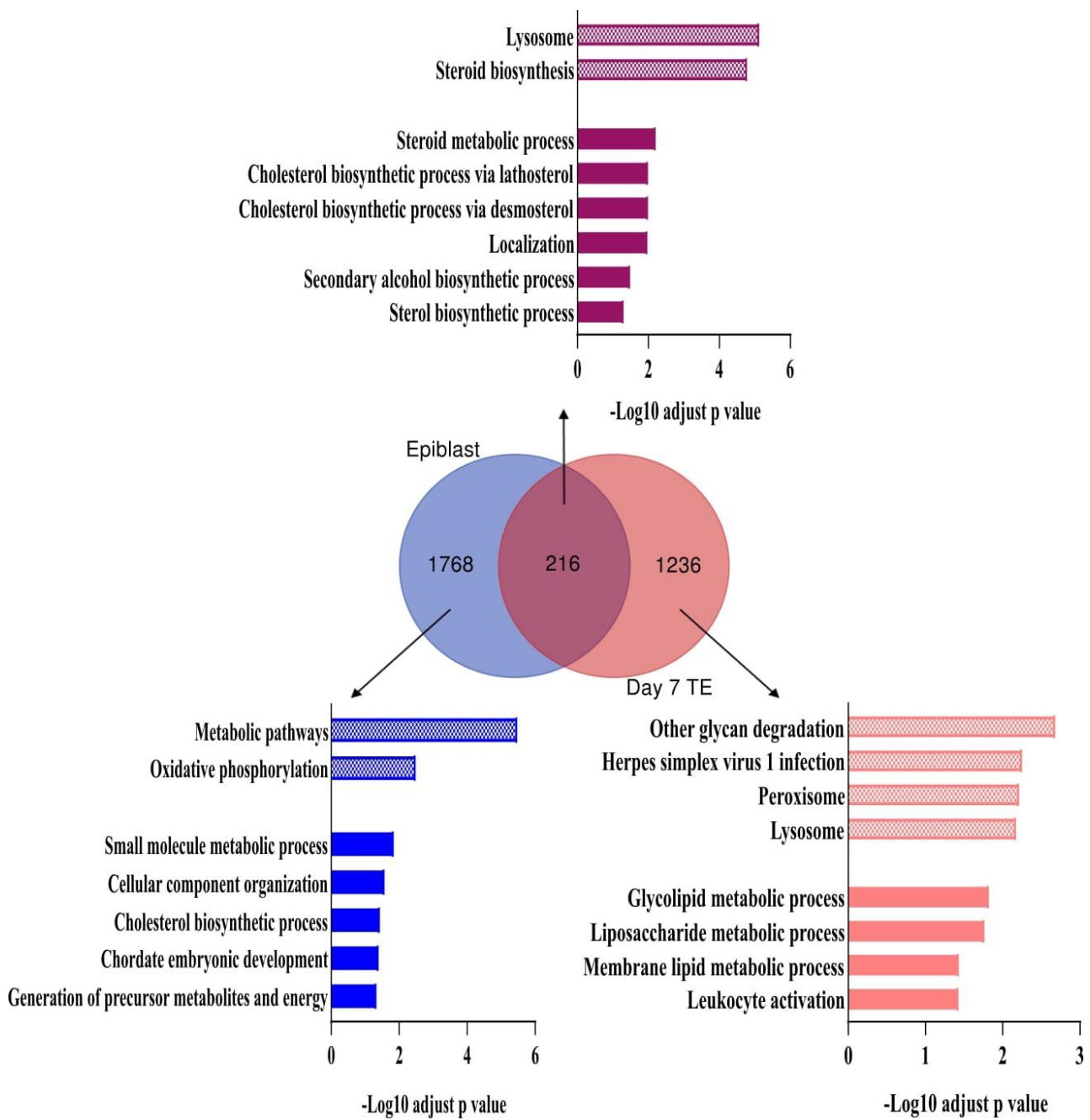

**Fig.6**. **Biological processes enriched in *ACE2*- and *TMPRSS2*-positive cells in epiblast and day 7 trophectodermal cells of developing human blastocysts.** (A) Venn diagram showing numbers of differentially abundant genes unique and common in *ACE2*- and *TMPRSS2*- positive cells vs *ACE2*- and *TMPRSS2*-negative cells in day 7 trophectodermal (D7 TE) cells and epiblast cells of human embryos. Biological processes enriched in the differential abundant genes of the epiblast (B) and day 7 TE (C). The Y axis indicates the enriched biological processes and X axis is –Log10 adjusted P value. Hatched bar are terms of the KEGG pathways, the solid bars represent the biological processes. Data were extracted from single-cell RNA-Seq of the epiblast of developing human embryos (Stirparo et al., 2018) . Data for day 7 (D7) blastocyst trophectodermal cells was extracted from (Petropoulos et al., 2016)

**Discussion**

The results of the present study demonstrate that early human embryos express coronavirus entry receptors and S protein proteases. The embryonic cells also expressed the genes for proteins that are involved in viral endocytosis and replication.

Considering the scale at which the SARS-CoV-2 virus has spread globally and the fact that a proportion of individuals harbouring the virus are asymptomatic, it is likely that some of the infected individuals may have conceived or are trying to conceive during the duration of the pandemic. Further, with nearly 1 in 6 couples facing infertility, many would resort to *in vitro* fertilization (IVF) techniques for achieving biological parenthood during the pandemic. In both these scenarios, it is imperative to understand whether the developing embryo is at risk for SARS-CoV-2 infection. To address this question, we analysed single-cell RNA-Seq datasets of developing human embryos for SCARFs in zygotes to hatched blastocysts. Our results reveal that gametes, zygotes, and 4-cell embryos express *ACE2* and *BSG* along with *CTSL* but not *TMPRSS2*. While the levels of *ACE2* decline in the compact morula, the expression of *ACE2* and the S protein processing enzymes increases in the ICM and epiblast of the blastocyst embryos. Along with the cells of the embryo proper, we also detected abundant expression of these genes in the trophectodermal lineage of the blastocysts. Although ACE2 is essential for SARS-CoV-2 infection, the processing of the viral spike protein by the membrane-bound serine protease TMPRSS2 promotes viral infectivity. We observed that even though more than 80% of cells of the ICM express *ACE2,* none of these cells express *TMPRSS2*. However, in the epiblast (but not the primitive endoderm) and the trophectodermal cells of the blastocysts, a significant proportion of cells co-express *ACE2* and *TMPRSS2*, indicating that the early embryonic cells may be permissible to viral entry by the canonical ACE2 mode of entry.

Beyond ACE2, a study has shown that the extracellular matrix metalloproteinase enhancer, CD147, encoded by the gene *BSG* binds to both SARS-CoV-2 and SARS-CoV, promoting viral entry independent of ACE2 and TMPRSS2 (Chen et al., 2005; Wang et al., 2020). We observed that *BSG* transcripts were abundantly expressed in all the cells of the developing embryos from the zygote stage to blastocyst stage including the trophectodermal lineage. With regard to S protein priming, cathepsins are a class of endosomal proteases, and of these, cathepsin L encoded by *CTSL* is required for endosomal cleavage of SARS-CoV and SARS-CoV-2 spike proteins (Hoffmann et al., 2020). Interestingly, along with *BSG*, *CTSL* was co-expressed in most of the cells of the developing embryo. In a subset of cells of the blastocyst cell stage, embryos also co-expressed *ACE2* and *BSG*. Presently, it is unclear if CD147-mediated viral entry requires cathepsin L or TMPRSS2; the fact that both these proteins are co-expressed in most trophectodermal cells as well as epiblast cells imply that both canonical and non-canonical modes of SARS-CoV-2 entry may be operative in developing embryos.

Two other members of the coronavirus family that are highly infectious and can cause significant mortality are MERS and HCoV-229E. Unlike SARS-CoV and SARS-CoV-2, MERS and HCoV-229E utilize DPP4 and ANPEP as receptors to infect the human host cells (Li, 2015). Interestingly, we found that *DPP4* and *ANPEP* transcripts are not expressed in the zygotes until the morula stage, but both receptors are expressed in almost all the cells of the ICM, epiblast, and early endoderm. Together our data for the first time suggest that early human embryos could be permissive to SARS-CoV-2 and other coronaviruses; embryos at the blastocyst stage may be most susceptible to infection. It will be of interest to study the protein expression of these genes in these cell types and experimentally determine the effects of SARS-CoV-2 (and other coronaviruses) on blastocyst health.

During development, the cells of the embryo proper are enveloped by a layer of trophectodermal cells that form the placenta. The placenta is a key determinant of the health of the developing fetus and protects it from maternal infections. We and others have shown that the trophoblast cells of the human placenta in the first to third trimester express SCARFs, suggesting that the placenta may be permissive to SARS-CoV-2 infection (Ashray et al., 2020; Singh et al., 2020b). Extending these findings, herein we show that the trophectodermal cells that are the precursors of the placenta abundantly express the SCARFs as early as day 5 when they make their first morphologic appearance. Furthermore, we also observed that unlike in the case of the placenta, where 5-15% of the trophoblasts express *ACE2* and *TMPRSS2* (Ashray et al., 2020), almost all the cells of the trophectoderm co-express both these factors. Similarly, almost all the trophectodermal cells of the blastocysts co-transcribe the non-canonical receptor CD147 and the S protein protease CTSL along with other coronavirus receptors ANPEP and DPP4. These results indicate that the trophectoderm of early human embryos may be highly permissive to multiple coronaviruses.

Once the virus binds to the receptor and the spike protein undergoes cleavage, SARS-CoV-2 releases its content into the host cells and initiates replication. At this point of the viral infectious cycle, most enveloped viruses recruit the ESCRT machinery of the host cell. The viral structural protein engages the tumour susceptibility gene protein (TSG101) in ESCRT-I, which subsequently delivers ESCRT-III to sites of viral budding where membrane scission releases new viral particles (Votteler and Sundquist, 2013). Several proteins are involved in the ESCRT pathway, and blocking any of these enzymes can prevent viral endocytosis (Wang et al., 2017; Mazzon and Marsh, 2019). Whether coronaviruses utilize ESCRT for endocytosis and budding is unknown but there is some evidence for the SARS-CoV-2 proteins interacting with human proteins of ESCRT pathway (Gordon et al., 2020; Maaroufi, 2020). Thus, we postulated that for SARS-CoV-2 to be pathogenic to human embryos, the early embryonic cell must possess the components of the ESCRT machinery for viral endocytosis and release. Interestingly, we observed that genes for many components of the ESCRT machinery are not expressed in zygote and morula stages, but a surge in the expression of these genes occurs in the cells of the blastocyst including the ICM, epiblast, and primitive endoderm and most abundantly in the cells of the trophectoderm. These results indicate that the human blastocysts expressing SARS-CoV-2 receptors not only have the necessary machinery for viral attachment onto the cell surface but also readily express the infrastructure to facilitate endocytosis and viral budding.

Once the virus enters the cells, the next step in the infectious cycle is replication, which requires interaction of the viral proteins with host proteins (de Wilde et al., 2018). Our results reveal that as compared to the zygote, 4- to 8-cell stages, and morula stage embryos, the cells of the blastocyst show a surge in mRNA expression of most host proteins involved in viral replication. Almost all of the proteins involved in viral replication were abundantly expressed in the ICM, epiblast, and primitive endoderm with the highest expression in the trophectoderm. Together our data show that early developing human embryos have the necessary machinery to facilitate viral replication.

Our analysis, so far, suggests that several genes involved in endocytosis and viral replication are expressed in the cells of developing embryos. However, it remains imperative to determine if all or a subset of the cells express these genes. Since most *ACE2*- and *TMPRSS2*-positive cells were present in the epiblast and the trophectoderm, we analysed the data at single-cell resolution and observed that irrespective of the expression of *ACE2* and *TMPRSS2*, cells of the epiblast and trophectoderm have transcripts of genes involved in viral endocytosis and replication, albeit with varying degrees of expression. These genes were also uniformly expressed in cells of the early ICM and the primitive endoderm (not shown). These results imply that along with the trophectoderm and epiblast, the other cells of the blastocyst could be susceptible to SARS-CoV-2 infection.

Once the viral RNA enters the host cells, it translates several of its non-structural proteins, which in turn physically interact with various host proteins to regulate host cell machinery. There are 332 human proteins that physically interact with SARS-CoV-2 (Gordon et al., 2020). These proteins have a wide range of functional roles in host cellular processes such as DNA replication, vesicle trafficking, lipid modification, RNA processing and regulation, nuclear transport machinery, cytoskeletal organization, mitochondrial functions, and extracellular matrix modelling (Gordon et al., 2020). We observed that mRNA of genes encoding for the SARS-CoV-2 interacting proteins are expressed in most cells of the ICM, primitive endoderm, and the epiblast; the mRNA levels of these genes were also high in the cells of the trophectodermal lineage. Together our observations imply that in human embryos, a high proportion of genes representing protein products that interact with SARS-CoV-2 are expressed in the cells of the embryo proper and even more so in the trophectoderm.

In developing embryos, the health of the cells of the epiblast and trophectoderm are crucial as these cells subsequently undergo gastrulation and placentation, respectively. Any damage to these cells, physically or functionally, may lead to embryo lethality or organ dysfunctions in later development. To understand the molecular characteristics of the SARS-CoV-2 permissive cells, we analysed the gene signature of the *ACE2+TMPRSS2+* epiblast and day 7 trophectodermal cells. We identified 1985 genes in the epiblast that were differentially expressed between the *ACE2-* and *TMPRSS2-* positive cells and *ACE2-* and *TMPRSS2*-negative cells; 1452 genes were differentially abundant in the trophectodermal lineage. Unique gene signatures of *ACE2*-positive cells have also been reported for other tissues (Seow et al., 2020; Xu et al., 2020; Ziegler et al., 2020). Amongst the overabundant genes in the *ACE2+TMPRSS2+* epiblast and trophectoderm cells were cathepsins (including CTSL) and members of the TMPRSS family, some of which are required for SARS-CoV-2 S protein cleavage (Hoffmann et al., 2020; Shang et al., 2020a; Zang et al., 2020). Beyond the proteases, mRNA levels of several other proteins such as TROP*2* (also known as *TACSTD2*) were also detected in higher abundance in the *ACE2+TMPRSS2+* cells of the epiblast and trophectoderm. In the adult human liver, the highest expression of *ACE2* and *TMPRSS2* are detected in *TROP2*-positive cells, and viral infection of the *TROP2+* progenitors may be a cause of hepatic failure in patients with COVID-19 (Seow et al., 2020). These observations imply that the *ACE2+TMPRSS2+* cells could have a unique profile of genes that could aid in viral infectivity and pathogenesis of COVID-19.

Single-cell RNA-Seq has revealed that in developing human embryos, the inner cell mass and trophectoderm segregate on day 5 of development, the trophectodermal lineage progressively gets specified, and by day 7 two distinct sub-populations of trophectoderm become evident (Petropoulos et al., 2016). In contrast, the epiblast retains pluripotency, and transcriptomic hallmarks of self-renewal are observed well until day 7 (Singh et al., 2019). Despite such developmental divergence, we observed that the *ACE2+TMPRSS2+* cells of the epiblast and trophectoderm shared a unique gene signature. The biological processes associated with the common genes involved sterol, alcohol, and cholesterol metabolism, and the steroid biosynthesis KEGG pathway was enriched. Host cell fatty acid synthesis machinery plays a crucial role in viral genome replication and virion production, and accumulation of sterols allows for production of secluded membranes for viral replication, which shields viral nucleic acids from immune surveillance (Mayer et al., 2019). In the context of coronaviruses, a fine-tuned host lipid profile is essential to achieve optimal viral replication (Yan et al., 2019). Along with lipid metabolism, the common genes in the *ACE2+TMPRSS2+* cells of the epiblast and trophectoderm are also enriched in lysosomes and lysosomal proteases. Endosome-lysosome fusion profoundly promotes infection, and lysosomal proteases are an important determinant of the species and tissue tropism of coronaviruses (Burkard et al., 2014; Zheng et al., 2018). Furthermore, genes in endocytosis and lysosome pathways are highly enriched in cells of the lungs of patients with COVID-19 (Xiong et al., 2020) and lysosomotropic drugs like hydroxychloroquine and others have potential for anti-SARS-CoV-2 activity (Gordon et al., 2020; Sarma et al., 2020). Thus, it appears that the *ACE2+TMPRSS2+* cell subsets in the developing embryo have a milieu conducive for viremia.

The *ACE2+TMPRSS2+* cells of the trophectoderm and epiblast also had unique gene signatures. In the trophectoderm pathways of herpes simplex viral infection, glycan modifications and peroxisomes were enriched. Glycans are essential components of viruses including SARS-CoV-2 and play a key role in their pathogenesis (Fanibunda et al., 2011; Watanabe et al., 2020). Peroxisomes are multifunctional organelles with roles in cellular metabolism and signalling; some virus-host interactions rewire host peroxisomes to support viral replication and spread (Lazarow, 2011). The role of host glycan modifications and peroxisome machinery in SARS-CoV-2 pathogenesis is yet to be understood; it is tempting to propose that since these pathways are enriched in the *ACE2+TMPRSS2+* cells, it will be worth exploring these processes for identifying novel drugs for SARS-CoV-2. In the epiblast, the unique genes in *ACE2+TMPRSS2+* cells, as expected, had roles in cellular organization for embryonic development, which also included enriched genes in oxidative phosphorylation. A higher expression of genes involved in metabolic pathways including oxidative phosphorylation is reported in neutrophils of COVID-19 patients (Gardinassi et al., 2020; Singh et al., 2020a). Together the results imply a possible role of mitochondrial activity in SARS-CoV-2 infection.

Thus, our analyses reveal that the *ACE2-* and *TMPRSS2*-positive cells in the epiblast and trophectoderm are enriched for genes involved with lipid metabolic pathways and lysosomes; the trophectoderm cells specifically are enriched for genes involved in the peroxisome while the epiblast cells are enriched for genes involved in mitochondrial functions. Since all these processes are utilized by viruses (including coronaviruses), it is tempting to propose that the coronaviruses have adapted to optimize access to the host cells that are highly permissive for viral propagation, and in different tissues, multiple pathways could be utilized for their spread.

Viral infections in pregnancy are associated with adverse outcomes such as miscarriages, stillbirths, and congenital abnormalities (Adams Waldorf *et al.*, 2018). Preimplantation embryos are considered to be protected because the zona pellucida acts as a barrier that reduces the susceptibility of embryonic cells to viral infection. However, reduced survival of zona-intact mouse embryos is observed upon exposure to Zika viruses (Tan et al., 2019; Block et al., 2020), suggesting that the zona pellucida may be more vulnerable than it is believed to be. In addition, the Zika virus has been shown to infect the trophectoderm and ICM of preimplantation embryos that express the Zika receptors, and exposure of early preimplantation embryos to Zika viruses compromises fetal brain development (Tan et al., 2019). Extrapolating from these observations and the fact that cells of the embryo proper and the trophectoderm are potentially permissive to viral entry, it is plausible that SARS-CoV-2 and other coronaviruses may also have embryo-toxic effects. Indeed, *ACE2* and *TMPRSS2*, which are highly expressed in the developing trophectoderm, are also expressed in the human placental cells, and virions are detected in the trophoblast cells of the placenta from a pregnant women with COVID-19 (Ashray et al., 2020; Hosier et al., 2020; Vivanti et al., 2020). Further, placental damage is observed in pregnancies complicated by SARS-CoV-2 infection (Bahadur et al., 2020; Verma et al., 2020). While it will be impossible to generate data on the long-term consequences of SARS-CoV-2 infection in human embryos, clinicians and patients seeking fertility options via in vitro fertilization specifically must be aware of the potential negative impact of SARS-CoV-2 on early developing embryos.

Overall, this study provides a roadmap for future studies on embryo-toxic effects of SARS-CoV-2 and other coronaviruses. While this is the first in-depth survey of SCARF expression in developing human embryos, it is limited by the constraints and shortcomings of scRNA-seq. These include under-representation of rare cell types, isolation biases, and statistical cutoffs. Furthermore, RNA expression levels do not always accurately reflect protein abundance; therefore, our observations must be corroborated for protein expression *in situ*. Also, SCARF expression may be modulated by co-infection of SARS-CoV-2 and other pathogens, resulting in completely diverse consequences.

Nevertheless, our study provides valuable baseline data on SCARFs in developing human embryos, which may prove to be a useful resource in the context of not just SARS-CoV-2 but also other viral infections in general. This knowledge will be useful to clinics performing IVF procedures during the COVID-19 pandemic in order to develop evidence-based guidelines for patient management.

**Conflict of interest**

We have no conflict of interest.

**Author contributions**

DM conceived the study and analysed the results. SC and KC took the lead role, executed the study, and analysed the data. AS, AH, RS, and AM carried out the literature survey and collected the information. NS, AB, DS, and NA helped in data assembly, analysis, and manuscript preparation. AB, DS and KC prepared the figures. All the authors agreed to the final version of the manuscript.

**Funding**

No specific funding was received for this study.

**Acknowledgments**

We express our gratitude to the Director (ICMR-NIRRH) Dr. Smita Mahale for constant guidance and encouragement. The following publicly available datasets were used in this study (E-MTAB-3929 and supplementary table 5 in (Stirparo et al., 2018) and we acknowledge the authors for making them available for research use. DM lab is funded by grants from Indian Council of Medical Research (ICMR), Govt. of India. SC is thankful to the Department of Health Research (DHR), Govt. of India for the Young Scientist fellowship. KC is thankful for the Nehru-Fulbright Scholarship. AH and NS are thankful to ICMR-RA and ICMR-SRF respectively. RS is thankful to DST-Inspire fellowship, AM and DS are thankful to UGC for research fellowship. NA and AB are thankful to DST-SERB project fellowship. The manuscript bears the NIRRH ID RA/894/04-2020.

**Contribution to the field**

This study is the first in-depth survey of coronavirus receptors and its associated factors in developing human embryos. We show that human embryonic and the trophectoderm cells (which develop in to the placenta) express the necessary factors that can aid in viral entry, replication and dissemination. Overall, this study provides evidence for potential embryo-toxic effects of SARS-CoV-2 and other coronaviruses. Clinically, this knowledge will be useful in developing guidelines for conducting *in vitro* fertilization and other assisted reproduction techniques.

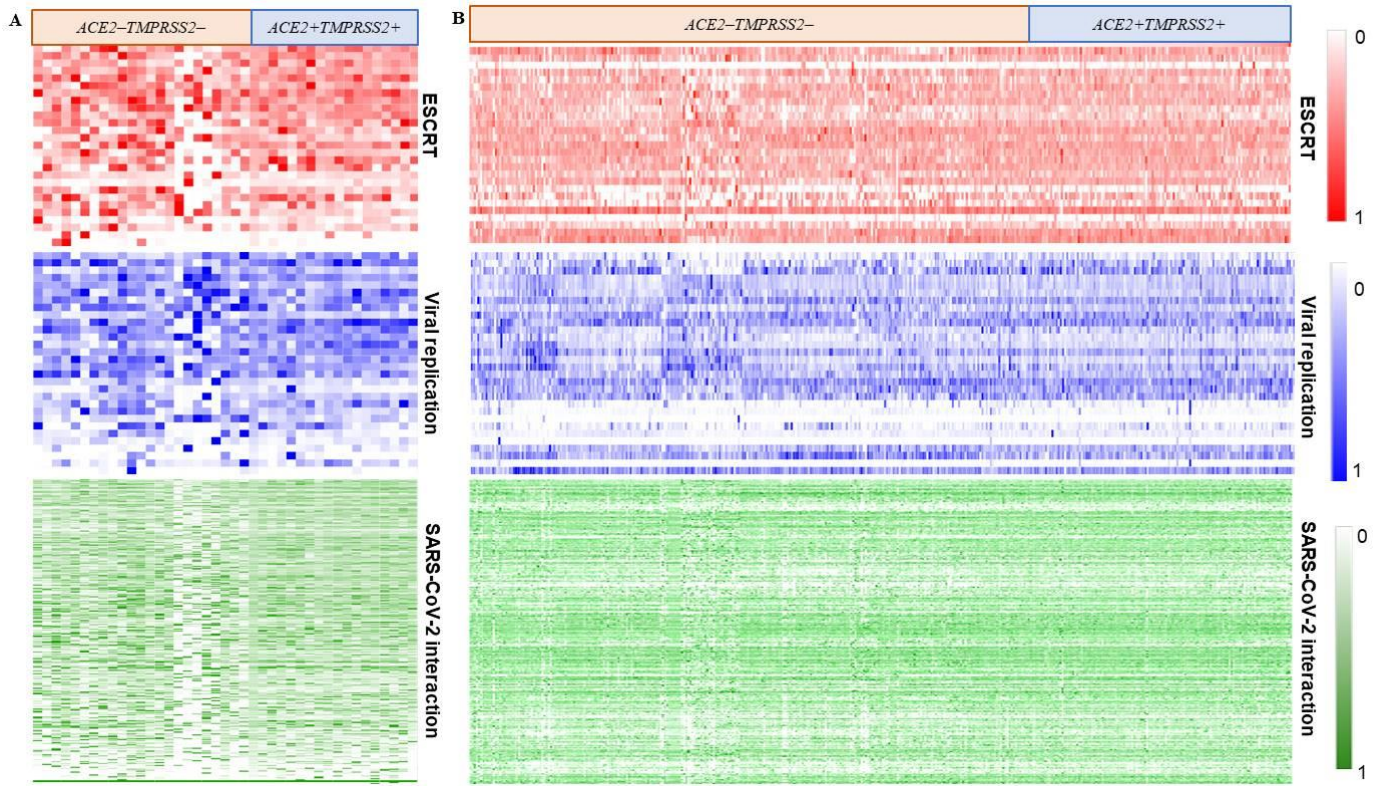

**Supplementary Fig 1**: mRNA levels of Endosomal Complexes Required for Transport (ESCRT), viral replication and SARS-CoV-2 interactions in single cells of epiblast (A) and trophectoderm (B). . In all heat maps, each row depicts a gene and each column depicts a single cell. The *ACE2+TMPRSS2+* are cells co-expressing both these genes and *ACE2-TMPRSS2-* are cells express neither. Data is presented in relative color scale. Single-cell RNA-Seq of epiblasts was extracted from (Stirparo et al., 2018) and trophectoderm cells from (Petropoulos et al., 2016).

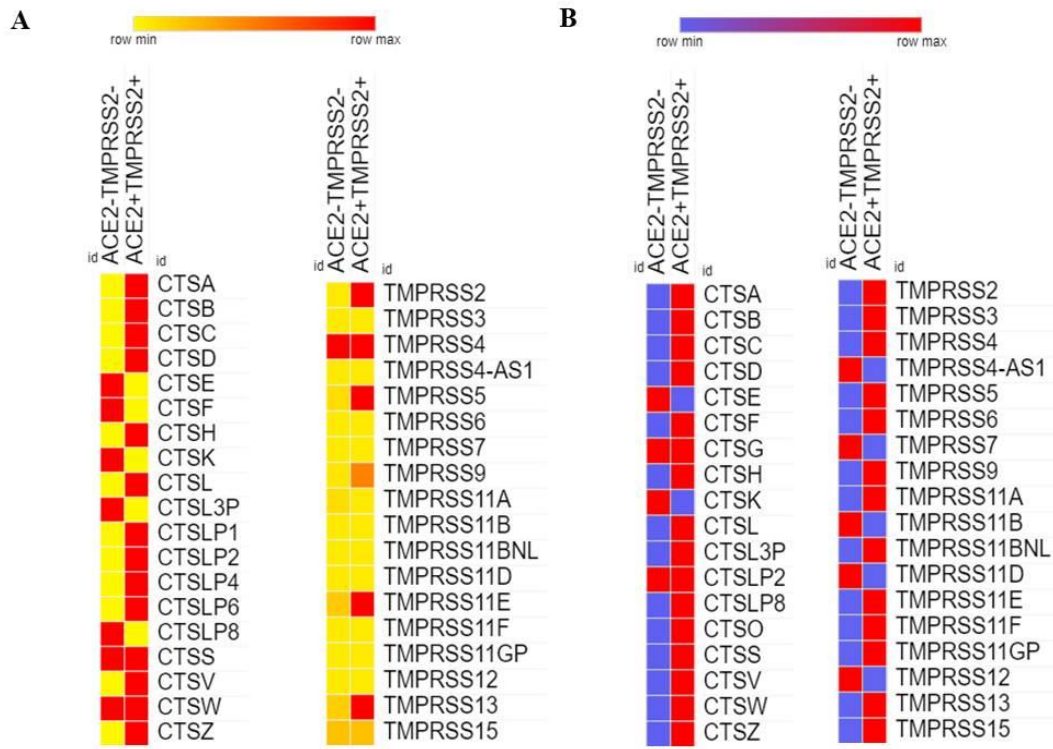

**Supplementary Fig 2: Heat maps for expression of Cathepsin (CTS) family and TMPRSS family genes in Epiblast (A) and Day 7 trophectoderm (B).** .Each row depicts a gene and each column is pseudo-bulk data of single cell. Data is presented in relative color scale. Single-cell RNA-Seq of epiblasts was extracted from (Stirparo et al., 2018) and trophectoderm cells from (Petropoulos et al., 2016).